\documentclass[10 pt,slashbox]{article}
\usepackage{indentfirst,mathrsfs}
\usepackage{amsfonts}
\usepackage{amsmath,amsthm,amssymb}
\usepackage{slashbox}
\usepackage{color}

\usepackage{graphicx}
\newtheorem{Theorem}{Theorem}
\newtheorem{Corollary}{Corollary}
\newtheorem{Definition}{Definition}
\newtheorem{Example}{Example}

\newtheorem{Lemma}{Lemma}
\newcommand{\F}{\mathbb{F}}
\newcommand{\Z}{\mathbb{Z}}

\begin{document}

\title{On the Systematic Constructions of Rotation Symmetric Bent  Functions with  Any  Possible  Algebraic Degrees}
\author{ Sihong Su and Xiaohu Tang
\thanks{The authors are with the Information Security
and National Computing Grid Laboratory, Southwest Jiaotong University, Chengdu,
 610031, China (e-mail: sush@henu.edu.cn, xhutang@swjtu.edu.cn).
 Sihong Su is also with the School of Mathematics and Statistics, Henan University, Kaifeng, 475004, China.}}
\date{}

\maketitle

\begin{abstract}

In the literature, few constructions of $n$-variable rotation
symmetric bent functions have been presented, which either  have
restriction on $n$ or have algebraic degree no more than $4$.  In
this paper, for any even integer $n=2m\ge2$, a first systemic
construction of $n$-variable rotation symmetric bent functions, with
any possible algebraic degrees ranging from $2$ to $m$, is proposed.
\end{abstract}


\section{Introduction}

Boolean bent functions were introduced by Rothaus in 1976
\cite{Rot76}. Let $\F_2$ be the finite field with two elements,
$n>0$ be a positive integer, and $\F_2^n$ be the $n$-dimensional
vectorspace over $\F_2$. An $n$-variable Boolean function from
$\F_2^n$ to $\F_2$ is \textit{bent} if it has maximal Hamming
distance to the set of affine Boolean functions. Boolean bent
functions have attracted much attention due to their important
applications in cryptography  \cite{Cant03,Charp05}, coding
theory and sequence design \cite{Lem82,Mac77,Ols82}.

Boolean functions that are invariant under the action of cyclic
rotation on the inputs are called rotation symmetric  functions \cite{Pie99}.
Such class of Boolean functions is of great interest since they
need less space to be stored and allow faster computation of the Walsh transform.
Further, it has been experimentally
demonstrated that the class of rotation symmetric   functions is
extremely rich in terms of cryptographically significant Boolean
functions. In particular, they allowed obtaining Boolean functions in
odd numbers of variables beating the best known nonlinearities
\cite{Kav07}, and new bent functions (in even numbers of variables)
\cite{Carl14,Carl14seta,Dala09,Gao}.

Throughout this paper, for $n=2m$ we study the $n$-variable
 rotation symmetric bent functions.  To avoid confusion, we denote
  the sum over $\Z$ by $+$, and the sum over $\F_2$ by
$\oplus$. The quadratic Boolean function
 \begin{eqnarray}\label{Eqn_f_0}
  f_0(x_0,\cdots,x_{n-1})=\bigoplus\limits_{i=0}^{m-1} x_ix_{m+i}
 \end{eqnarray}
is the first class of rotation symmetric bent functions. According
to experimental results, St\v{a}nic\v{a} \textit{et al.} conjectured
that there is no homogeneous rotation symmetric bent function having
algebraic degree greater than 2 \cite{Stani08}. Since then, large
classes of homogeneous rotation symmetric  functions seem  to
support the conjecture since all of them do not contain
non-quadratic bent functions \cite{Meng10,Stanica08}.

As any theoretic advancement in this direction can be used to find
cryptographically significant functions on higher number of
variables, it was stated in \cite{Stani08} that any theoretic
construction  of rotation symmetric bent functions with algebraic
degree larger than 2 is an interesting problem. In the literature,
the main method of constructing new rotation symmetric bent
functions  is to  modify $f_0(x)$ in \eqref{Eqn_f_0}
\cite{Carl14seta,Gao}. Up to now, only few constructions of rotation
symmetric bent functions are known, whose  algebraic degrees are all
no more than  $4$. In  \cite{Gao}, for $n=2m$, Gao \textit{et al.}
proved the cubic rotation  symmetric
 function
\begin{eqnarray*}
 f_t(x_0,\cdots,x_{n-1})=\bigoplus\limits_{i=0}^{m-1} x_ix_{m+i}\oplus\bigoplus\limits_{i=0}^{n-1}
 \big(x_ix_{t+i}x_{m+i}\oplus x_ix_{t+i}\big)
\end{eqnarray*}
is a rotation symmetric bent function  if and only if
$\frac{m}{\mathrm{gcd}(m,t)}$ is odd, where $1\le t\le m-1$
and the subscript of $x$  is modulo $n$. This is the first theoretical
construction of rotation symmetric bent functions with algebraic
degree lager than $2$.  Recently, another  $n$-variable cubic rotation
symmetric bent function
\begin{eqnarray*}
f(x_0,\cdots,x_{n-1})=\bigoplus\limits_{i=0}^{m-1}x_ix_{m+i}\oplus\bigoplus\limits_{i=0}^{n-1}
x_ix_{r+i}x_{2r+i}\oplus
\bigoplus\limits_{i=0}^{2r-1}x_ix_{2r+i}x_{4r+i},
\end{eqnarray*}
where $n=2m=6r$, was presented in \cite{Carl14seta}. Later on, an infinite class of $n$-variable rotation symmetric
bent functions with algebraic degree 4, where $n=2m$ but not
divisible by 4, was constructed from two known semi-bent rotation
symmetric  functions in $m$ variables with complementary Walsh
supports \cite{Carl14}.% by using the indirect sum given in \cite{Car10}. %In fact,
%although the algebraic normal form of the  functions  in
%\cite{Carl14} is too complex to express, it also contain the term
%$\bigoplus\limits_{i=0}^{m-1}x_ix_{m+i}$.

In this paper, we present a simple but generic construction of
$n$-variable rotation symmetric bent functions still by the
modification of  the quadratic rotation symmetric bent function
$f_0$  in \eqref{Eqn_f_0}. Unlike the previous modifications, our
construction can provide $n$-variable rotation symmetric bent
functions for any even integer $n$. Most notably, the proposed
$n$-variable rotation symmetric bent functions can have any possible
algebraic degree ranging from 2 to $n/2$. To the best of our
knowledge, it is the first time to construct rotation symmetric bent
functions of algebraic degree larger than $4$ when $n\ge 10$.

The rest of this paper is organized as follows. In Section 2,
some basic notations and definitions of Boolean functions,
rotation symmetric bent functions in particular, are reviewed. In Section 3,
a generic construction of  $n$-variable rotation
symmetric bent functions is proposed by modifying the support  of the quadratic
rotation symmetric bent function $f_0$  in \eqref{Eqn_f_0}. In Section 4,
  a flexible construction of  $n$-variable rotation
symmetric bent functions   with any given  algebraic degree from 2 to $n/2$ is presented.
Finally,  Section 5 concludes this paper.

\section{Preliminaries}

Given a vector $x=(x_0,\cdots,x_{n-1})\in
\F_2^n$,  define its support  as the set $\textrm{supp}(x)=\{0\le
i< n \,|\,x_i=1 \}$, and its Hamming weight $\mathrm{wt}(x)$ as the
cardinality of its support, i.e.,
$\mathrm{wt}(x)=|\mathrm{supp}(x)|$.

In this paper, for simplicity, we do not distinguish the vector
$x=(x_0,\cdots,x_{n-1})\in \F_2^n$ and the integer
$\sum_{i=0}^{n-1} x_i 2^i\in\{0,\cdots,2^n-1\}$ if the
context is clear, since they are one-to-one corresponding. %For
%example, when $n=2$, then $(0,0)=0$, $(1,0)=1$, $(0,1)=2$,
%$(1,1)=3$.
 For any two vectors $x=(x_0,\cdots,x_{n-1})\in \F_2^n$
and $y=(y_0,\cdots,y_{n-1})\in \F_2^n$, if $x_i\le y_i$ for all $0\le i<
n $, then we say that $y$ covers  $x$ and denote it by $y\succeq x$. According to
Lucas formula, we have
\begin{eqnarray}\label{Eqn_Lucas}
{y\choose x}=1\pmod2\Longleftrightarrow y\succeq x.
\end{eqnarray}

 Let $x=(x_0,\cdots,x_{n-1} )\in \mathbb{F}_2^n$.  For two integers
 $l\geq 0$ and $0\le i< n $,  define
the left $l$-cyclic shift version of vector $x$ as
$\rho_l^n(x)=\big(\rho_l^n(x_0),\cdots,\rho_l^n
(x_{n-1})\big)$ by
\begin{eqnarray*}
\rho_l^n(x_i)=x_{i+l},
\end{eqnarray*}
where the subscript of $x$  is modulo $n$.
An orbit generated by a vector
$x\in\F_2^n$ is defined as
\begin{eqnarray}\label{Eqn_Ox}
O_n(x)=\big\{\rho_0^n(x),\cdots,\rho_{n-1}^n(x)\big\}.
\end{eqnarray}
In other words, each orbit consists of all cyclic shifts of one vector in
  $\F_2^n$. Naturally, an orbit   in $\F_2^n$
can be represented by its representative element which is the
lexicographically first element belonging to the orbit. The set of
the representative elements of all the  orbits in $\F_2^n$ is
denoted by $\mathbf{R}_n$.  For example,  $\mathbf{R}_4=\{(0, 0, 0, 0), (1, 0, 0, 0), (1, 1, 0, 0), (1, 0, 1, 0),
(1, 1, 1, 0), (1, 1, 1, 1)\}.$

 An $n$-variable Boolean
function    is a mapping from $\F_2^n$ into $\F_2$. We denote by
$\mathcal{B}_n$ the set of all the $n$-variable Boolean functions. A
basic representation of a function $f\in\mathcal{B}_n$ is by the
output of its truth table, i.e., a binary vector of length
$2^n$, as
\begin{eqnarray*}\label{Eqn_TT}
f = [f(0),  \cdots, f(2^n-1)].
\end{eqnarray*}
The support of $f$ is defined as $\mathrm{supp}(f) = \{x
\in\F_2^n\,|\, f(x) = 1\}$ and $f$ is also said to be the characteristic function of the set $\mathrm{supp}(f)$. The Hamming weight of $f$ is the
cardinality of $\mathrm{supp}(f)$, i.e.,
$\mathrm{wt}(f)=|\mathrm{supp}(f)|$. It is easy to see that
$\mathrm{supp}(f_0)=O_4(1,0,1,0)\cup O_4(1,1,1,0)$ and
$\mathrm{wt}(f_0)=6$ for $f_0$ in \eqref{Eqn_f_0} when $n=4$.
% By
%convenience, define $\mathrm{zeros}(f)=\{x\in \F_2^n\,|\,f(x)=0\}$
%as well.
 %We say that a   function $f\in\mathcal{B}_n$ is balanced
% if its truth table contains an equal number of $1$'s and $0$'s,
% i.e., if its Hamming weight equals $2^{n-1}$.

The most usual representation of a Boolean function $f\in
\mathcal{B}_n$ is the algebraic normal form (ANF) as
\begin{eqnarray}\label{Eqn_ANF_Def}
f(x)=\bigoplus\limits_{\alpha\in
\F_2^n}c_{\alpha}x^\alpha,~c_{\alpha}\in \F_2,
\end{eqnarray}
where $c_{\alpha}$ is   the coefficient of the term
$x^\alpha=x_0^{\alpha_0} \cdots x_{n-1}^{\alpha_{n-1}}$  for
$x=(x_0,\cdots,x_{n-1} )$ and
$\alpha=(\alpha_0,\cdots,\alpha_{n-1})$ in $\F_2^n$. The algebraic degrees of
the term $x^\alpha$ and the  Boolean function $f$ in
\eqref{Eqn_ANF_Def} are respectively defined as
$\mathrm{deg}(x^\alpha)=\mathrm{wt}(\alpha)$ and
\begin{eqnarray*}\label{Eqn_deg}
\mathrm{deg}(f)=\max\{\mathrm{wt}(\alpha)\,|\,c_{\alpha}=1,\alpha\in
\F_2^n\}.
\end{eqnarray*}
Specifically, the Boolean functions of degree at most $1$ are called affine
functions; the function $f$ in \eqref{Eqn_ANF_Def} is called to be
homogeneous if all the terms with nonzero coefficients in $f$ have
the same algebraic degree.

\begin{Definition} \label{Def_rota} For a function $f\in\mathcal{B}_n$, if
$f(\rho_l^n(x))=f(x)$ holds  for all inputs $x\in
\mathbb{F}_2^n$ and integers $1\leq l \leq n-1$, then   $f$ is called a rotation symmetric function. That is,
rotation symmetric  functions are invariant under cyclic
rotation on inputs.
\end{Definition}

The Walsh transform of an $n$-variable Boolean function $f$ is an
integer-valued function on $\F_2^n$, whose value at
$\alpha\in\F_2^n$ is defined as
\begin{eqnarray}\label{Walsh}
W_f(\alpha)=\sum\limits_{x\in\F_2^n}(-1)^{f(x)\oplus \alpha\cdot x}
\end{eqnarray}
where $\alpha\cdot x=\alpha_0x_0\oplus\cdots\oplus \alpha_{n-1}
x_{n-1}$ is the usual inner product of
$\alpha=(\alpha_0,\cdots,\alpha_{n-1})$ and
$x=(x_0,\cdots,x_{n-1})$. The nonlinearity of a function
$f\in\mathcal{B}_{n}$ is given by
 $nl(f)=2^{n-1}-{1\over
   2}\max\limits_{\alpha\in\F_2^n}|W_f(\alpha)|$.

\begin{Definition}\label{Def_bent} A Boolean function $f:\F_2^n\rightarrow\F_2$
 is said to be bent if $W_f(\alpha)=\pm2^{\frac{n}{2}}$  for all $\alpha\in\F_2^n$.
 \end{Definition}

%If the algebraic degree of an
%$n$-variable bent function $f$ is maximal, then we say $f$ has
%optimal algebraic degree.

Obviously, an $n$-variable Boolean
function is bent only if $n$ is even.  In addition, it is well known that the algebraic degree of an
$n$-variable bent function is no more than $m$ for $n=2m\ge4$, while
the algebraic degree of a $2$-variable bent function is 2.

The following result will be used in the computation of the values
of the Walsh transform later.

\begin{Lemma}\label{Lem_omegax}
Let $a, b$ be two vectors over $ \F_2^n$. Then,
\begin{eqnarray*}\label{Eqn_alpbetx}
\sum\limits_{x\in\F_2^n}(-1)^{x\cdot(x\oplus a \oplus b)}= \left\{
  \begin{array}{ll}
    2^n, & a=\overline{b} \\
    0, & \mathrm{otherwise} \\
  \end{array}
\right.
\end{eqnarray*}
where $\overline{b}=(b_0\oplus 1,\cdots,b_{n-1}\oplus 1)$ for
$b=(b_0,\cdots,b_{n-1})$.
\end{Lemma}

\begin{proof}
For any two vectors $a, b\in \F_2^n$,
 \begin{eqnarray*}
&& \sum\limits_{x\in\F_2^n}(-1)^{x \cdot(x\oplus a\oplus b)}\\
&=& \sum\limits_{x\in\F_2^n}(-1)^{x\cdot((1,\cdots,1)\oplus a\oplus b)}\\
%&=&\left\{
%  \begin{array}{ll}
%    2^n, & \alpha\oplus\beta\oplus1=0 \\
%    0, & \mathrm{otherwise} \\
%  \end{array}
%\right.\\
&=&\left\{
  \begin{array}{ll}
    2^n, & a=\overline{b} \\
    0, & \mathrm{otherwise} \\
  \end{array}
\right.
\end{eqnarray*}
where the first identity holds since $x\cdot x=\mathrm{wt}(x)=
x\cdot(1,\cdots,1)$, and the second identity holds by the fact that
$\sum\limits_{y\in\F_2^n}(-1)^{\lambda\cdot y}=2^n$ if $\lambda=0$
and $\sum\limits_{y\in\F_2^n}(-1)^{\lambda\cdot y}=0$ if
$\lambda\in\F_2^n\setminus\{0\}$.
\end{proof}

From now on, we always assume $n=2m\ge2$. For a  vector
$x=(x_0,\cdots,x_{n-1})\in\F_2^n$, we always denote
$x'=(x_0,\cdots,x_{m-1})$, $x''=(x_{m},\cdots,x_{n-1})$, $x'\cdot
x''=x_0x_{m}\oplus \cdots\oplus x_{m-1}x_{n-1}$, and
$x'*x''=x_0x_{m}+\cdots+ x_{m-1}x_{n-1}$. Obviously,  $x'*x''=0$ if and
only if $x_ix_{m+i}=0$ for $0\le i\le m-1$.
%$\mathrm{supp}(x')\cap\mathrm{supp}(x'')=\emptyset$.
 For simplicity, we use the notation $x'+x''$ for
$x'\oplus x''$ satisfying $x'*x''=0$. This is to say,  when $x'+x''$ is used in the rest of
this paper, it always implies
$x'*x''=0$.

\section{A generic construction of rotation symmetric bent  functions}

In this section, we present a generic construction of rotation
symmetric bent functions by modifying the support  of  $f_0(x)$ in
\eqref{Eqn_f_0}.

  Given a subset $T\subseteq\F_2^n$,
define
 an $n$-variable Boolean function as
\begin{eqnarray}\label{Eqn_fun}
f(x)= \left\{
  \begin{array}{ll}
    f_0(x) \oplus 1, &x\in T\\
     f_0(x), & \mathrm{otherwise} \\
  \end{array}
\right.
\end{eqnarray}
 where  $f_0$  is given in \eqref{Eqn_f_0}.
In order to construct an $n$-variable rotation symmetric bent
function $f$ in \eqref{Eqn_fun}, it is crucial to choose a proper
subset $T$ of $\F_2^n$.

Firstly, we give a sufficient and necessary condition of $T$ such
that
 $f$ in \eqref{Eqn_fun} is a rotation symmetric function.

\begin{Lemma}\label{Lem_Torbit}
The $n$-variable Boolean function $f$ in \eqref{Eqn_fun} is a
rotation symmetric function if and only if $O_n(x)\subseteq T$ for
all $x\in T$.
\end{Lemma}

\begin{proof} Recall that  $f_0$ is a rotation symmetric function. Let
$\chi_T(\cdot)$ be  the characteristic function of $T$. Then, by
Definition \ref{Def_rota}, $f$ is  a rotation symmetric function if
and only if $\chi_T=f\oplus f_0$ is also a rotation symmetric
function, which is equivalent to $\rho_l^n(x)\in T$ for all $x\in T$
and $1\le l<n$. This completes the proof by the definition of $O_n(x)$
in \eqref{Eqn_Ox}.
\end{proof}

Secondly, we study a sufficient condition of $T$ such that $f$ is a
bent function.

\begin{Lemma}\label{Lem_bentfun}
For any subset $\Gamma\subseteq\F_2^m$, if the subset
\begin{eqnarray}\label{Eqn_T-condition}
 T=\bigcup\limits_{\gamma\in
\Gamma}\big\{x\in\F_2^n\,|\,x'\in\F_2^m,x''=x'\oplus
 \gamma\big\},
\end{eqnarray}
  then the $n$-variable Boolean function $f$ in
\eqref{Eqn_fun} is a   bent function.
 \end{Lemma}

\begin{proof} Substituting $f$ %in \eqref{Eqn_fun}
 to the definition of the Walsh transform in \eqref{Walsh}, we have
 \begin{eqnarray*}
W_{f}(\alpha )
&=&\sum\limits_{x \in \F_2^n\setminus T}(-1)^{f_0(x)\oplus\alpha\cdot x }+\sum\limits_{x\in T}(-1)^{f_0(x)\oplus1\oplus\alpha\cdot x }\\
&=&\sum\limits_{x\in\F_2^{n}}(-1)^{x'\cdot x''\oplus\alpha'\cdot x'\oplus\alpha''\cdot x''}-2\sum\limits_{x\in T}(-1)^{x'\cdot x''\oplus\alpha'\cdot x'\oplus\alpha''\cdot x''}\\
&=&\sum\limits_{x''\in\F_2^m}(-1)^{\alpha'' \cdot x''}\sum\limits_{x'\in\F_2^m}(-1)^{(x''\oplus\alpha')\cdot x'}-2\sum\limits_{\gamma\in \Gamma}\sum\limits_{x'\in \F_2^m}(-1)^{x'\cdot (x'\oplus \gamma)\oplus\alpha'\cdot x'\oplus\alpha''\cdot (x'\oplus \gamma)}\\
&=&(-1)^{\alpha'\cdot\alpha''}2^m-2\sum\limits_{\gamma\in \Gamma}(-1)^{\alpha''\cdot \gamma}\sum\limits_{x'\in \F_2^m}(-1)^{x'\cdot (x'\oplus \gamma\oplus\alpha'\oplus\alpha'')}\\
&=&\left\{
                      \begin{array}{ll}
                      (-1)^{1+\alpha'\cdot\alpha''}2^m, & \mathrm{if~}\alpha'\oplus\alpha''=\overline{\gamma} ~\mathrm{for ~a} ~\gamma\in \Gamma\\
                        (-1)^{\alpha'\cdot\alpha''}2^m, & \mathrm{otherwise} \\
                      \end{array}
                    \right.
\end{eqnarray*}
where the fourth identity comes from the fact that
$\sum\limits_{y\in\F_2^m}(-1)^{\lambda\cdot y}=2^m$ if $\lambda=0$
and $\sum\limits_{y\in\F_2^m}(-1)^{\lambda\cdot y}=0$ if
$\lambda\in\F_2^m\setminus\{0\}$, the last identity follows from
Lemma \ref{Lem_omegax}.
\end{proof}

The following result is immediate from Lemmas \ref{Lem_Torbit} and
\ref{Lem_bentfun}.

\begin{Theorem}\label{Thm_cons_Talpha}
The $n$-variable Boolean function defined in \eqref{Eqn_fun} is a
rotation symmetric bent  function if the subset $T\subseteq\F_2^n$ satisfies
\eqref{Eqn_T-condition} and $O_m(\gamma)\subseteq \Gamma$ for all
$\gamma\in \Gamma$.
\end{Theorem}

\begin{proof}  According to
\eqref{Eqn_T-condition}, $x=(x_0,\cdots,x_{n-1})\in T$ if and only
if  $x_{i+m}=x_i\oplus\gamma_{i}$, i.e.,
$x_{l+i+m}=x_{l+i}\oplus\gamma_{l+i}$,  for all $0\le i<n$ and $1\le
l<n$, where $\gamma=(\gamma_0,\cdots,\gamma_{m-1})\in \Gamma$. %, and
%the subscript of $x$ $($resp. $\gamma$$)$   is taken as modulo $n$
%$($resp. $m$$)$ if it is larger than $n$$($resp. $m$$)$.
 Then,
given $1\le l<n$,  $\rho_l^n(x)=(x_{l},\cdots,x_{l+n-1}) \in T$ if
and only if $\rho_l^m(\gamma)=(\gamma_l,\cdots,\gamma_{l+m-1})\in
\Gamma$, i.e., $O_n(x)\subseteq T$ if and only if
$O_m(\gamma)\subseteq \Gamma$. Hence, $f$  in \eqref{Eqn_fun} is a
rotation symmetric bent function by Lemmas \ref{Lem_Torbit} and
\ref{Lem_bentfun}.
\end{proof}

In what follows, we investigate the ANF of the function proposed in
Theorem \ref{Thm_cons_Talpha}. To do so, it is sufficient to
determine the ANF of the characteristic function $\chi_T$.

Given a $\gamma\in \mathbf{R}_m$, define
\begin{eqnarray}\label{Eqn_Talpha}
T_{\gamma}=\bigcup\limits_{\delta \in
O_m(\gamma)}\big\{x\in\F_2^n\,\big|\, x'\in\F_2^m,x''=x'\oplus
\delta  \big\}.
      \end{eqnarray}
It is easy to see that the subset $T_{\gamma}$ of $\F_2^{n}$ in
\eqref{Eqn_Talpha} satisfies:
\begin{itemize}
  \item [P1] $ T_{\alpha}\cap T_{\beta} =\emptyset$ if $\alpha\not=\beta$, where $\alpha,\beta\in \mathbf{R}_m$.
%  \item [P2] $\bigcup\limits_{\gamma\in \mathbf{R}_m}T_{\gamma}=\F_2^n$, which gives a partition of $\F_2^n$.
\end{itemize}

\begin{Example}

If  $n=4$, the vector sets $T_{\gamma}$, $\gamma\in \mathbf{R}_2$,
are given in Table \ref{Tab_Tgamma}.

\begin{table*}[!ht]
\begin{center}
\caption{The vector sets $T_{\gamma}$ for $\gamma\in \mathbf{R}_2$
}\label{Tab_Tgamma}
\begin{tabular}{|c|c|c| c|c|c|c| c}
 \hline   $\gamma$ &  $T_{\gamma}$ \\
\hline $(0,0)$  & $O_4(0,0,0,0)\cup O_4(1,0,1,0)\cup O_4(1,1,1,1)$   \\
\hline $(1,0)$  &  $O_4(1,0,0,0)\cup O_4(1,1,1,0) $  \\
\hline $(1,1)$  & $O_4(1,1,0,0)$  \\
\hline
 \end{tabular}
\end{center}
\end{table*}

\end{Example}

Since $O_m(\gamma)\subseteq \Gamma$ for all $\gamma\in \Gamma$, then
we can write $\Gamma$ as $\Gamma=\bigcup\limits_{\gamma\in
\mathbf{R}_m\cap \Gamma} O_m(\gamma)$  and then
\begin{eqnarray*}
T=\bigcup\limits_{\gamma\in \mathbf{R}_m\cap \Gamma} T_{\gamma}.
\end{eqnarray*}
By P1, we have
\begin{eqnarray}\label{Eqn_Chi_Tx}
\chi_T(x)=\bigoplus\limits_{\gamma\in \mathbf{R}_m\cap \Gamma}
\chi_{T_{\gamma}}(x)
\end{eqnarray}
where $\chi_T(\cdot)$ is  the characteristic
function of $T$. Therefore, we study the ANF of the function $\chi_{T_{\gamma}}$
firstly.

\begin{Lemma}\label{Lem_cons_Talpha'}
The ANF of the $n$-variable characteristic function of $T_\gamma$ in
\eqref{Eqn_Talpha} is
\begin{eqnarray}
\chi_{T_\gamma}(x)&=&\bigoplus\limits_{\delta\in O_m(\gamma)}
 \bigoplus\limits_{
 \begin{subarray}{c}
    \beta'+ \beta''\succeq \delta\\
    %\beta'* \beta''=0
 \end{subarray}
}x^{\beta}\label{Eqn_funxalpha-1}\\
&=&\bigoplus\limits_{
\begin{subarray}{c}
  \delta\in\mathbf{R}_m \\
 \delta\succeq\gamma
\end{subarray}
}c_{\delta}\bigoplus\limits_{
 \begin{subarray}{c}
    \beta'+ \beta''\in O_m(\delta)\\
    %\beta'* \beta''=0
 \end{subarray}
}x^{\beta}\label{Eqn_funxalpha-2}
\end{eqnarray}
where $c_{\delta}\in\{0,1\}$ with $c_{\gamma}=1$.
\end{Lemma}

\begin{proof}  First of all, we prove \eqref{Eqn_funxalpha-1} in two special cases: $\gamma=\mathbf{0}_m$ and $\gamma=\mathbf{1}_m$ where
$\mathbf{0}_m$ and $\mathbf{1}_m$ respectively denote the all-zero
and all-one vector of length $m$.

According to the definition of characteristic function, we have
\begin{eqnarray}\label{Eqn_ANF_T1}
\chi_{T_{\mathbf{1}_m}}(x)&=& \bigoplus
\limits_{(b_0,\cdots,b_{m-1})\in\F_2^m}\prod\limits_{i=0}^{m-1}(x_{i}\oplus
b_{i}\oplus1)(x_{m+i}\oplus
 b_{i}) \nonumber\\
&=& \prod\limits_{i=0}^{m-1}\Big( \bigoplus\limits_{b_i\in\{0,1\}} (x_{i}\oplus b_{i}\oplus1)(x_{m+i}\oplus b_{i})\Big) \nonumber\\
&=&\prod\limits_{i=0}^{m-1} (x_i\oplus x_{m+i})\nonumber\\
%&=&\bigoplus\limits_{
%\begin{subarray}{c}
%    0\le i< m \\
%  (\beta_i,\beta_{m+i})\in\{(1,0),(0,1)\}
%  \end{subarray}
%  }\prod\limits_{i=0}^{m-1} x_i^{\beta_i} x_{m+i}^{\beta_{m+i}}\nonumber\\
&=&\bigoplus\limits_{
\begin{subarray}{c}
    0\le i< m \\
   \beta_i,\beta_{m+i} \in\F_2, \beta_i+\beta_{m+i}=1
\end{subarray}
}\prod\limits_{i=0}^{m-1} x_i^{\beta_i} x_{m+i}^{\beta_{m+i}}%\nonumber\\
%&=& \bigoplus\limits_{
% \begin{subarray}{c}
%   \beta'+\beta'' =\mathbf{1}_m \\
%   %\beta'*\beta''=0
% \end{subarray}
%}x^{\beta}
\end{eqnarray}
and
\begin{eqnarray}\label{Eqn_ANF_T0}
\chi_{T_{\mathbf{0}_m}}(x)&=& \bigoplus
\limits_{(b_0,\cdots,b_{m-1})\in\F_2^m}\prod\limits_{i=0}^{m-1}(x_{i}\oplus
b_{i}\oplus1)(x_{m+i}\oplus
 b_{i}\oplus1) \nonumber\\
&=& \prod\limits_{i=0}^{m-1}\Big( \bigoplus\limits_{b_i\in\{0,1\}} (x_{i}\oplus b_{i}\oplus1)(x_{m+i}\oplus b_{i}\oplus 1)\Big) \nonumber\\
&=&\prod\limits_{i=0}^{m-1} (x_i\oplus x_{m+i}\oplus 1)\nonumber\\
&=&\bigoplus\limits_{
\begin{subarray}{c}
   0\le i< m \\%(\beta_i,\beta_{m+i})\in\{(0,0),(1,0),(0,1)\} \\
 \beta_i,\beta_{m+i} \in\F_2, \beta_i+\beta_{m+i}\in\{0,1\}
\end{subarray}
}\prod\limits_{i=0}^{m-1} x_i^{\beta_i} x_{m+i}^{\beta_{m+i}}%\nonumber\\
%&=&  \bigoplus\limits_{
% \begin{subarray}{c}
%    \beta'*\beta''=0
% \end{subarray}
%}x^{\beta}
\end{eqnarray}

Then, based on \eqref{Eqn_ANF_T1} and \eqref{Eqn_ANF_T0}, we are
able to get
\begin{eqnarray*}
&&   \chi_{T_{\gamma}}(x)\\
 &=&\bigoplus_{
       \begin{subarray}{c}
         (b_0,\cdots,b_{m-1})\in\F_2^m  \\
         \delta=(\delta_0,\cdots,\delta_{m-1})\in O_m(\gamma)
       \end{subarray}
      }\prod_{i=0}^{m-1}(x_i\oplus b_i\oplus1)(x_{m+i}\oplus b_i\oplus \delta_i\oplus1)\\
   % &&(by~\eqref{Eqn_Talpha})\\
 &=&\bigoplus_{
    \begin{subarray}{c}
       (b_0,\cdots,b_{m-1})\in\F_2^m  \\
         \delta\in O_m(\gamma)
    \end{subarray}
     }\prod_{i\in\mathrm{supp}(\delta)}(x_i\oplus b_i\oplus 1)(x_{m+i}\oplus b_i)
    \prod_{i\in\mathrm{zeros}(\delta)}(x_i\oplus b_i\oplus 1)(x_{m+i}\oplus b_i\oplus 1)\\
%&=& \bigoplus_{\delta\in O_m(\gamma)}\Bigg(\bigoplus_{
%    \begin{subarray}{c}
%         i\in\mathrm{supp}(\delta) \\
%         (\beta_i,\beta_{m+i})\in\{(1,0),(0,1)\}
%    \end{subarray}}\prod_{
%         i\in\mathrm{supp}(\delta)
%    } x_i^{\beta_i} x_{m+i}^{\beta_{m+i}}\Bigg)\Bigg(\bigoplus_{
%    \begin{subarray}{c}
%         i\in\mathrm{zeros}(\delta) \\
%         (\beta_i,\beta_{m+i})\in\{(0,0),(1,0),(0,1)\}
%    \end{subarray}
%    }\prod_{i\in\mathrm{zeros}(\delta)} x_i^{\beta_i} x_{m+i}^{\beta_{m+i}} \Bigg)\\
&=& \bigoplus_{\delta\in O_m(\gamma)}\Bigg(\bigoplus_{
    \begin{subarray}{c}
         i\in\mathrm{supp}(\delta) \\
          \beta_i,\beta_{m+i}\in\F_2, \beta_i+\beta_{m+i}=1
    \end{subarray}
    } \prod_{i\in\mathrm{supp}(\delta)}x_i^{\beta_i} x_{m+i}^{\beta_{m+i}}\Bigg)
    \Bigg(\bigoplus_{
    \begin{subarray}{c}
         i\in\mathrm{zeros}(\delta)\\
          \beta_i,\beta_{m+i}\in\F_2, \beta_i+\beta_{m+i}\in\{0,1\}
    \end{subarray}
    }\prod_{i\in\mathrm{zeros}(\delta)} x_i^{\beta_i} x_{m+i}^{\beta_{m+i}} \Bigg)\\
&=& \bigoplus_{\delta\in O_m(\gamma)}\Bigg(\bigoplus_{
    \begin{subarray}{c}
          \beta_i,\beta_{m+i}\in\F_2, 1\le i\le m  \\
          \beta_i+\beta_{m+i}=1,i\in\mathrm{supp}(\delta)\\
          \beta_i+\beta_{m+i}\in\{0,1\},i\in\mathrm{zeros}(\delta)
    \end{subarray}    }x^{\beta}\Bigg)\\
&=&  \bigoplus_{\begin{subarray}{c}
              \delta\in O_m(\gamma)
              \end{subarray}
      }\bigoplus_{
    \begin{subarray}{c}
               \beta'+\beta''\succeq\delta
    \end{subarray}    }x^{\beta}\\
       \end{eqnarray*}
where $\mathrm{zeros}(x)=\{0\le i< n\,|\,x_i= 0\}$ for
$x=(x_0,\cdots,x_{n-1})\in\F_2^n$ and the third identity follows
from \eqref{Eqn_ANF_T1} and \eqref{Eqn_ANF_T0}.

Next we prove \eqref{Eqn_funxalpha-2}.
Note the fact that
$x\succeq \rho_k^m(\gamma)$ if and only if $\rho_{m-k}^m(x)\succeq
\gamma$ for any $x\in \F_2^m$ and $0\le k<m$. Therefore, by the definition of $O_n(x)$
in \eqref{Eqn_Ox} we have
$\{x\in\F_2^m\,|\,x\succeq \delta,\, \delta\in
O_m(\gamma)\}=\{x\in\F_2^m\,|\,x\in O_m(\delta),\,\delta \succeq
\gamma,\,\delta\in \mathbf{R}_m \}$. Then, we can rewrite
\eqref{Eqn_funxalpha-1} as
\begin{eqnarray}\label{Eqn_funxalpha-3}
\chi_{T_\gamma}(x)=\bigoplus\limits_{
\begin{subarray}{c}
  \delta\in\mathbf{R}_m \\
 \delta\succeq\gamma
\end{subarray}
}
 \bigoplus\limits_{
 \begin{subarray}{c}
    \beta'+ \beta''\in O_m(\delta)\\
    %\beta'* \beta''=0
 \end{subarray}
}c_{\beta}x^{\beta}
\end{eqnarray}
where
$c_{\beta}=c_{\beta}'\,(\bmod\,2)$ and $c_{\beta}'$  is the number
of the term $x^{\beta}$ that appears in the right hand side of
\eqref{Eqn_funxalpha-1}, i.e.,
$c_{\beta}'=|\{\delta|\beta'+\beta''\succeq \delta, \delta\in
O_m(\gamma)\}|$. Still by the
above fact,  we have that $c_{\beta}'$ is a constant for
all $\beta'+\beta''\in O_m(\delta)$, which is denoted by $c_{\delta}'$
for convenience, clearly $c_{\gamma}'=1$.
Then we arrive at \eqref{Eqn_funxalpha-2} from \eqref{Eqn_funxalpha-3}
where $c_{\delta}=c_{\delta}'\,(\bmod\,2)$ and  $c_{\gamma}=c_{\gamma}'\,(\bmod\,2)=1$.
\end{proof}

\begin{Example}

When $n=4$, the ANFs of  $\bigoplus\limits_{
 \begin{subarray}{c}
    \beta'+ \beta''\succeq \delta\\
    %\beta'* \beta''=0
 \end{subarray}
}x^{\beta}$ and $\bigoplus\limits_{\beta'+\beta''\in
O_m(\delta)}x^{\beta}$, $\delta\in \F_2^2$, are given in Table
\ref{Tab_TgammaANF}.

\begin{table*}[!ht]
\begin{center}
\caption{The ANFs  of  $\bigoplus\limits_{
    \beta'+ \beta''\succeq \delta
}x^{\beta}$, $\bigoplus\limits_{\beta'+\beta''\in
O_m(\delta)}x^{\beta}$, $\delta\in \F_2^2$ }\label{Tab_TgammaANF}
\begin{tabular}{|c|c|c| c|c|c|c| c}
 \hline   $\delta$ &  $\bigoplus\limits_{\beta'+ \beta''\succeq \delta}x^{\beta}$ &$\bigoplus\limits_{\beta'+\beta''\in  O_m(\delta)}x^{\beta}$\\
\hline $(0,0)$  & $1\oplus \bigoplus\limits_{i=0}^3(x_i\oplus x_ix_{i+1})$ & $1$  \\
\hline $(1,0)$  &  $  x_0\oplus x_2\oplus\bigoplus\limits_{i=0}^3x_ix_{i+1}  $ & $\bigoplus\limits_{i=0}^{3}x_i$ \\
\hline $(0,1)$  &  $ x_1\oplus x_3\oplus\bigoplus\limits_{i=0}^3x_ix_{i+1} $ & $\bigoplus\limits_{i=0}^{3}x_i$ \\
\hline $(1,1)$  & $\bigoplus\limits_{i=0}^3x_ix_{i+1} $ & $\bigoplus\limits_{i=0}^{3} x_ix_{i+1} $ \\
\hline
 \end{tabular}
\end{center}
\end{table*}

\end{Example}

Applying Lemma \ref{Lem_cons_Talpha'} to \eqref{Eqn_Chi_Tx}, we
have

\begin{Theorem}\label{Thm_Chi}
For the rotation symmetric bent  function given in Theorem
\ref{Thm_cons_Talpha}, its ANF is
\begin{eqnarray*}
f_0(x)\oplus\bigoplus\limits_{\gamma\in \mathbf{R}_m\cap
\Gamma}\bigoplus\limits_{\delta\in O_m(\gamma)}
 \bigoplus\limits_{
 \begin{subarray}{c}
    \beta'+ \beta''\succeq \delta\\
    %\beta'* \beta''=0
 \end{subarray}
}x^{\beta}.
\end{eqnarray*}
\end{Theorem}

As mentioned before, $n/2$ is the maximal  algebraic degree of the
$n$-variable bent  function, which is usually of particular
interest.

\begin{Corollary}\label{Cro_Maxdegree}
For the rotation symmetric bent  function given in
\ref{Thm_cons_Talpha}, the algebraic degree arrives at the maximal
value $n/2$ if and only if the size $|\Gamma|$ of $\Gamma$ is odd.
\end{Corollary}

\section{Rotation symmetric bent  functions of any possible algebraic degree}

In this section,  we  study a flexible construction of $n$-variable
rotation symmetric bent  functions
%containing terms
of any prescribed algebraic degree from
$2$ to $n/2$. We begin from a very useful
linear combination of the  $n$-variable  characteristic functions
$\chi_{T_{\gamma}}(x)$ in \eqref{Eqn_funxalpha-2}.

\begin{Lemma}\label{Lem_ANF}
For each $\delta\in \mathbf{R}_m$, there exists an nonempty subset
$A_{\delta}\subseteq \mathbf{R}_m$ such that
\begin{eqnarray*}
\bigoplus_{\beta'+\beta''\in
O_m(\delta)}x^{\beta}=\bigoplus\limits_{\gamma\in
A_{\delta}}\chi_{T_{\gamma}}.
\end{eqnarray*}
\end{Lemma}

\begin{proof}

List all the vectors   in $\mathbf{R}_m$ according to the Hamming
weight firstly and the lexicographic order secondly  as
$$\mathbf{R}_m=\{\alpha_1,\cdots,\alpha_{|\mathbf{R}_m|}\}$$
i.e., $\alpha_i\not\succeq\alpha_j$ if $i<j$. Then, by \eqref{Eqn_funxalpha-2} we have
\begin{eqnarray*}
\big(\chi_{T_{\alpha_1}},\cdots,\chi_{T_{\alpha_{|\mathbf{R}_m|}}}\big)
=\Big(\bigoplus_{\beta'+\beta''\in O_m(\alpha_1)}x^{\beta},\cdots,
\bigoplus_{\beta'+\beta''\in
O_m(\alpha_{|\mathbf{R}_m|})}x^{\beta}\Big) \left(
  \begin{array}{ccccc}
    1 & 0 & \cdots & 0 &0\\
    *  & 1 & \ddots  &  &0 \\
   \vdots  & \ddots  & \ddots  & \ddots& \vdots \\
   * &  &   \ddots & 1  &0\\
     *  &  * &  \cdots  & *&1 \\
  \end{array}
\right)
\end{eqnarray*}
Since the matrix is a lower triangular matrix of full rank, each
$\bigoplus\limits_{\beta'+\beta''\in O_m(\delta)}x^{\beta}$ can be expressed as a
linear combination of
$\chi_{T_{\alpha_1}},\cdots,\chi_{T_{\alpha_{|\mathbf{R}_m|}}}$.
\end{proof}

Based on Lemma \ref{Lem_ANF}, we can construct $n$-variable
rotation symmetric bent  function $f$
with any algebraic degree $2\le \mathrm{deg}(f)\le n/2$.

\begin{Theorem}\label{Thm_bentccons}
For any element $\delta\in\mathbf{R}_m$ with $\mathrm{wt}(\delta)\ge2$% and $\delta\ne (1,0,\cdots,0)$
, the function
\begin{eqnarray*}\label{Eqn_Con_ANF}
 f_0(x)\oplus
 \Big(
 \bigoplus\limits_{
 \begin{subarray}{c}
    \beta'+ \beta''\in O_m(\delta)
 \end{subarray}
}x^{\beta}\Big)
\end{eqnarray*}
is a rotation symmetric bent function with algebraic degree $\mathrm{deg}(f)=\mathrm{wt}(\delta)$, where $f_0$ is given in
\eqref{Eqn_f_0}.
\end{Theorem}

\begin{proof}
The bent property of $f$ is a direct consequence of \eqref{Eqn_Chi_Tx}, Theorem \ref{Thm_Chi}, and Lemma \ref{Lem_ANF}. And
$\mathrm{deg}(f)=\mathrm{wt}(\delta)$ comes the fact that
$\mathrm{deg}(x^\beta)=\mathrm{wt}(\beta)=\mathrm{wt}(\delta)$ for all $ \beta'+ \beta''\in O_m(\delta)$.

\end{proof}

By means of Theorem \ref{Thm_bentccons}, we are able to construct more $n$-variable rotation symmetric bent  functions by flexibly assembling some
$\bigoplus\limits_{\beta'+ \beta''\in O_m(\delta)}x^{\beta}$.

\begin{Theorem}\label{Thm_bentccons-1}
For any nonempty subset $A\subseteq\mathbf{R}_m$, % and $A\ne \{(1,0,\cdots,0)\}$,
the function
\begin{eqnarray*}\label{Eqn_Con_ANF}
 f_0(x)\oplus
 \bigoplus\limits_{\delta\in A}\Big(
 \bigoplus\limits_{
 \begin{subarray}{c}
    \beta'+ \beta''\in O_m(\delta)
 \end{subarray}
}x^{\beta}\Big)
\end{eqnarray*}
is a rotation symmetric bent function, where $f_0$ is given in
\eqref{Eqn_f_0}.
\end{Theorem}

\begin{Example} In   \cite{Carl14seta}, for $n=2m=6r$,
Carlet et al. constructed an   $n$-variable cubic rotation symmetric
bent function  as
\begin{eqnarray*}
f(x)=f_0(x)\oplus
\bigoplus\limits_{i=0}^{n-1}x_ix_{r+i}x_{2r+i}\oplus
\bigoplus\limits_{i=0}^{2r-1}x_ix_{2r+i}x_{4r+i}
\end{eqnarray*}
where $f_0(x)$ is given by
\eqref{Eqn_f_0}. According to Theorem 1 in \cite{Carl14seta}, we know that
$f(x)$ can be rewritten as
\begin{eqnarray*}
f(x)=f_0(x)\oplus \bigoplus\limits_{
 \begin{subarray}{c}
    \beta'+ \beta''\in O_m(\delta) \\
   %\beta'*\beta''=0
 \end{subarray}
}x^{\beta}
\end{eqnarray*}
where  $\delta\in \mathbf{R}_m$ such that
$\mathrm{supp}(\delta)=\{0,r,2r\}$. In other words, the cubic
rotation symmetric bent function in \cite{Carl14seta} is a simple
case of our construction.

 \end{Example}

Finally, we demonstrate a class of rotation symmetric bent functions
by setting
\begin{eqnarray}\label{Eqn_Tim}
\Gamma_i^{(m)}&=&\bigcup\limits_{
     \gamma\in \mathbf{R}_m,\,
         \mathrm{wt}(\gamma)=i}T_{\gamma}
\end{eqnarray}
where $T_{\gamma}$ is given in \eqref{Eqn_Talpha} and $0\le i<m$.

\begin{Lemma}\label{Lem_rep}  The ANF of the $n$-variable characteristic function of  $\Gamma_i^{(m)}$ in \eqref{Eqn_Tim} can be
expressed as
\begin{eqnarray*}
\chi_{\Gamma_i^{(m)}}(x)=\bigoplus\limits_{j\succeq i}
\bigoplus\limits_{
 \begin{subarray}{c}
    \mathrm{wt}(\alpha)=j\\
   \alpha'*\alpha''=0
 \end{subarray}
}x^{\alpha}
\end{eqnarray*}
for all $0\le  i\le m$.
\end{Lemma}

\begin{proof}
By \eqref{Eqn_Chi_Tx}, \eqref{Eqn_funxalpha-1} and \eqref{Eqn_Tim}, we  have
\begin{eqnarray*}
\chi_{\Gamma_{i}^{(m)}}(x)&=&\bigoplus\limits_{
     \begin{subarray}{c}
          \gamma\in \mathbf{R}_m\\
         \mathrm{wt}(\gamma)=i
      \end{subarray}
      }\chi_{T_{\gamma}}(x)\\
&=&\bigoplus\limits_{
     \begin{subarray}{c}
          \gamma\in \mathbf{R}_m\\
         \mathrm{wt}(\gamma)=i
      \end{subarray}
      }\bigoplus\limits_{\delta\in O_m(\gamma)}
   \bigoplus\limits_{
   \begin{subarray}{c}
     \beta'+ \beta''\succeq \delta\\
     %\beta'* \beta''=0
   \end{subarray}
   }x^{\beta}\\
&=&\bigoplus\limits_{
         \mathrm{wt}(\delta)=i}
   \bigoplus\limits_{
   \begin{subarray}{c}
     \beta'+ \beta''\succeq \delta\\
     %\beta'* \beta''=0
   \end{subarray}
   }x^{\beta}\\
&=& \bigoplus\limits_{j=i}^{m}\Bigg[{j\choose i} \bigoplus\limits_{
     \begin{subarray}{c}
           \mathrm{wt}(\beta)=j \\
           \beta'*\beta''=0
     \end{subarray}}x^{\beta}\Bigg]\\
&=&\bigoplus\limits_{j\succeq i} \bigoplus\limits_{
 \begin{subarray}{c}
    \mathrm{wt}(\beta)=j\\
   \beta'*\beta''=0
 \end{subarray}
}x^{\beta}
\end{eqnarray*}
where the fourth identity holds since given any  vector
$\beta\in\F_2^n$ with $\mathrm{wt}(\beta)=j$ for $i\le j\le m$, satisfying $\beta'+\beta''\succeq\delta$, we have that
\begin{itemize}
  \item $\mathrm{wt}(\beta)=\mathrm{wt}(\beta'+\beta'')$;
  \item the number of distinct $\delta$ with $\mathrm{wt}(\delta)=i$ satisfying $\beta'+\beta''\succeq\delta$
    is $j\choose i$,
\end{itemize}
and the last identity   holds by   Lucas  formula in \eqref{Eqn_Lucas}.
\end{proof}

From Theorem \ref{Thm_cons_Talpha}, Lemma \ref{Lem_rep},  \eqref{Eqn_Chi_Tx} and \eqref{Eqn_Tim}, we obtain the following theorem.

\begin{Theorem}\label{Thm_bentccons-2}
For any set $\Gamma_i^{(m)}$ in \eqref{Eqn_Tim}, the function
$f_0(x)\oplus\bigoplus\limits_{j\succeq i} \bigoplus\limits_{
 \begin{subarray}{c}
    \mathrm{wt}(\alpha)=j\\
   \alpha'*\alpha''=0
 \end{subarray}
}x^{\alpha}$
is a rotation symmetric bent function with algebraic degree $\mathrm{deg}(f)=m$, where $f_0$ is given in
\eqref{Eqn_f_0}.
\end{Theorem}

\section{Conclusion}
In this paper, for $n=2m$,  we proposed a systematic method for
constructing  $n$-variable rotation symmetric bent functions with
any given possible algebraic degrees ranging from $2$ to $m$.

 % The total number of $8$-variable  rotation symmetric bent functions is
%  $4\times3776\approx2^{14}$, and the total number of $10$-variable  rotation symmetric bent functions
%is $4\times4697347604\approx2^{34}$ \cite{Dala09}.

\end{document}